\documentclass[aps,prd,reprint,twocolumn,groupedaddress,nofootinbib]{revtex4-2}

\usepackage[hidelinks]{hyperref}
\usepackage{graphicx}
\usepackage{dcolumn}
\usepackage{bm}
\usepackage{bm}
\usepackage{bbm}
\usepackage{amsmath}
\usepackage{etoolbox}
\usepackage{float}
\usepackage{mhchem}
\usepackage{gensymb}
\usepackage{xcolor}
\usepackage{cancel}
\usepackage{tikz}
\usepackage{pgfplots}
\usepackage{multirow}
\pgfplotsset{compat=1.18}
\usepackage{tikz}
\usepackage{tikz-feynman}
\tikzfeynmanset{compat=1.1.0}
\usetikzlibrary{positioning}
\usetikzlibrary{calc}
\tikzfeynmanset{compat=1.1.0}

\def\bea{\begin{eqnarray}}
\def\eea{\end{eqnarray}}
\def\be{\begin{equation}}
\def\ee{\end{equation}}

\definecolor{seagreen}{rgb}{0.18, 0.55, 0.34}

\usepackage{etoolbox}
\usepackage[normalem]{ulem}

\begin{document}


\title{Cosmic birefringence from a stiff axion field and its cosmological implications}

\author{Andrea Minotti}\email{andrea.minotti@unito.it}
\affiliation{Dipartimento di Fisica, Universit\`{a} di Torino, via P. Giuria 1, I--10125 Torino, Italy}
\affiliation{Istituto Nazionale di Fisica Nucleare, Sezione di Torino, via P. Giuria 1, I--10125 Torino, Italy}
\author{Yunzhi Wu}\email{yunzhi.wu@unito.it}
\affiliation{Dipartimento di Fisica, Universit\`{a} di Torino, via P. Giuria 1, I--10125 Torino, Italy}
\affiliation{Istituto Nazionale di Fisica Nucleare, Sezione di Torino, via P. Giuria 1, I--10125 Torino, Italy}
\author{Tiziano Zanzarella}\email{tiziano.zanzarella@unito.it}
\affiliation{Dipartimento di Fisica, Universit\`{a} di Torino, via P. Giuria 1, I--10125 Torino, Italy}
\affiliation{Istituto Nazionale di Fisica Nucleare, Sezione di Torino, via P. Giuria 1, I--10125 Torino, Italy}
\author{Nicolao Fornengo}\email{nicolao.fornengo@unito.it}
\affiliation{Dipartimento di Fisica, Universit\`{a} di Torino, via P. Giuria 1, I--10125 Torino, Italy}
\affiliation{Istituto Nazionale di Fisica Nucleare, Sezione di Torino, via P. Giuria 1, I--10125 Torino, Italy}
\author{Marco Regis}\email{marco.regis@unito.it}
\affiliation{Dipartimento di Fisica, Universit\`{a} di Torino, via P. Giuria 1, I--10125 Torino, Italy}
\affiliation{Istituto Nazionale di Fisica Nucleare, Sezione di Torino, via P. Giuria 1, I--10125 Torino, Italy}
\author{Marco Taoso}\email{marco.taoso@to.infn.it}
\affiliation{Istituto Nazionale di Fisica Nucleare, Sezione di Torino, via P. Giuria 1, I--10125 Torino, Italy}

\begin{abstract}
We propose a model in which the observed cosmic microwave background birefringence arises from the interaction of photons with a stiff ($w=1$) pseudo-scalar field that today contributes a subdominant fraction of the total energy density, distinct from both dark matter and dark energy. 
Big Bang Nucleosynthesis constrains the cosmological density of the stiff field, leading to a lower bound for its coupling with photons, $g_{\phi\gamma} > 8.9 \times 10^{-14}\,\mathrm{GeV}^{-1}$.
Photon conversion in cluster-hosted quasars sets an upper bound, $g_{\phi\gamma} < 5.0 \times 10^{-13}\,\mathrm{GeV}^{-1}$.
Therefore, there is a narrow window for the discovery of this cosmological component that can be explored in the forthcoming years with astrophysical axion searches.
Moreover, 
we show that this scenario leads to gravitational-wave signals that can be tested with planned experiments. 
\end{abstract}

\maketitle


\section{Introduction} \label{Sec:Introduction}
When a polarized electromagnetic wave travels in a region of space permeated by pseudo-scalar fields, often referred to as axion-like particles (ALPs), the plane of linear polarization rotates by an angle $\beta$ due to the axion coupling to electromagnetism that induces different propagation velocities for left- and right-circularly polarized photons. This effect is independent of the frequency of the electromagnetic wave and is known as cosmic birefringence (CB)~\cite{Carroll:1989vb,Harari:1992ea,Lue:1998mq,Fujita:2020ecn}.
There is now intriguing evidence for the isotropic cosmic birefringence of cosmic microwave background (CMB) photons coming from the experiments Planck~\cite{Diego-Palazuelos:2022dsq,Eskilt:2022cff,Sullivan:2025wnv,Ballardini:2025apf,Remazeilles:2025wzd} and Atacama
Cosmology Telescope (ACT)~\cite{ACT:2025fju,Diego-Palazuelos:2025dmh}.
When the two datasets are combined, the statistical significance reaches $\gtrsim 4\sigma$~\cite{ACT:2025fju,Diego-Palazuelos:2025dmh,Yin:2025fmj,Eskilt:2026imm}. 
Both axion dark matter and axion dark energy have been widely investigated as possible sources of cosmic birefringence (see e.g., the review in Ref.~\cite{Komatsu:2022nvu} and references therein), and ways of discriminating between the two have been discussed~\cite{Sherwin:2021vgb,Nakatsuka:2022epj,Lee:2022udm}.
In this work, we propose a new model for cosmic birefringence in which the birefringence angle of CMB photons is due to their interaction with a today-subdominant stiff ($w=1$) pseudo-scalar field that does not constitute neither dark matter nor dark energy. Tracing back the energy density of this field, we expect
that in the early universe there has been
an epoch in which the energy density was dominated by the stiff field, a so-called "kination" era. Recently, the possibility of an early kination era has been studied in the context of axiogenesis and gravitational-wave signals, see e.g.~\cite{Co:2019wyp,Co:2020jtv,Co:2021lkc,Gouttenoire:2021jhk,Gouttenoire:2021wzu}, including the possibility to provide an explanation for the stochastic gravitational-wave background evidenced by Pulsar Timing Array (PTA) collaborations~\cite{Harigaya:2023pmw}.

Our work aims at providing a new possible (and very predictive) explanation for cosmic birefringence that differs from the already well-studied DM and DE paradigms and that can, at the same time, lead to testable gravitational-wave signatures.

The paper is organized as follows. Sec.~\ref{Sec:Model} presents the equation of motion and the equation of state of the cosmological pseudo-scalar field. In Sec.~\ref{Sec:pheno}, we describe the phenomenology of the stiff axion associated with cosmic birefringence, Big Bang Nucleosynthesis (BBN), conversion into photons, and gravitational waves, determining the allowed range of the axion-photon coupling and of the duration of a kination era. 
Sec.~\ref{Sec:Conclusion} brings our conclusions. In Appendix~\ref{Sec:Nokination} we discuss a scenario without a kination epoch. 

\section{The model} \label{Sec:Model}

Let us consider a massless pseudo-scalar field $\phi$ coupled to the electromagnetic field $F_{\mu\nu}$ described by the Lagrangian:
\begin{equation}\label{eq:2.1}
    \mathcal{L} = \frac{1}{2}\partial_{\mu}\phi\partial^{\mu}\phi -\frac{1}{4}F_{\mu\nu}F^{\mu\nu} - \frac{g_{\phi\gamma}}{4}\phi\widetilde{F}^{\mu\nu}F_{\mu\nu},
\end{equation}
where
\begin{equation}
    \widetilde{F}^{\mu\nu} \equiv \frac{1}{2}\epsilon^{\mu \nu \rho \sigma} F_{\rho \sigma}
\end{equation}
and 
\begin{equation}
    - \frac{g_{\phi\gamma}}{4}\phi\widetilde{F}^{\mu\nu}F_{\mu\nu} = g_{\phi\gamma} \phi (\textbf{E}\cdot \textbf{B})\;,
\end{equation}
with $g_{\phi\gamma}$ being the effective coupling, $\epsilon^{\mu \nu \rho \sigma}$ the Levi-Civita symbol, $\textbf{E}$ the electric field and $\textbf{B}$ the magnetic field.
For a homogeneous scalar field, the equation of motion and the equation of state are:
\begin{equation}
    \ddot{\phi}+3H\dot{\phi}-g_{\phi\gamma} (\textbf{E}\cdot \textbf{B}) =0
\end{equation}
and 
\begin{equation}
    w_{\phi}=\frac{p_{\phi}}{\rho_{\phi}}=\frac{\dot{\phi}^2/2 + g_{\phi\gamma}\phi (\textbf{E}\cdot \textbf{B})}{\dot{\phi}^2/2 - g_{\phi\gamma} \phi (\textbf{E}\cdot \textbf{B})},
\end{equation}
respectively, with $p_{\phi}$ being the pressure and $\rho_{\phi}$ the density of the field.

At high energies, the primordial plasma behaves as a highly conducting medium, thereby screening any large-scale electric field.
We can therefore neglect terms proportional to $(\mathbf{E}\cdot\mathbf{B})$~\footnote{We checked that the backreaction of this term on the evolution of the scalar field is negligible and no primordial magnetic fields are significantly enhanced~\cite{Sharma:2026hzp}.}, so that
\bea
\ddot{\phi}+3H\dot{\phi}=0 , \quad\quad\quad w_{\phi}=1.
\eea
In this regime, the scalar field behaves as a stiff fluid. In the literature, an epoch during which the energy density of the Universe is dominated by a stiff fluid, with equation of state $w=1$, is commonly referred to as kination, see Ref.~\cite{Gouttenoire:2021jhk} for an overview.

In what follows, it will be useful to express the time derivative of the pseudoscalar field as a function of redshift. Using $dz/dt=-(1+z)H(z)$, the solution of the equation of motion is
\begin{equation}\label{eq:2.8}
\dot{\phi}(z)=\dot{\phi}_* \left( \frac{1+z}{1+z_*} \right)^3,
\end{equation}
where $\dot{\phi}_*$ denotes the time derivative of the pseudoscalar field at recombination, corresponding to $z_*\simeq 1100$.

\section{Phenomenology}\label{Sec:pheno}
\subsection{Cosmic birefringence}\label{Ssec:CB}

The coupling of the pseudo-scalar field to electromagnetism induces different propagation velocities for left- and right-circularly polarized photons, giving rise to a rotation of the polarization plane that can accumulate over cosmological distances. This phenomenon is known as cosmic birefringence~\cite{Carroll:1989vb,Harari:1992ea,Lue:1998mq,Fujita:2020ecn} and is usually studied in the context of axion DM and axion DE~\cite{Fedderke:2019ajk,Berbig:2024aee,Urena-Lopez:2025rad,Tada:2024znt,Lee:2025yvn,Nakagawa:2025ejs,Lin:2025gne,Barman:2025ryg,Minotti:2026ivu}.\\
Here we consider a (today) subdominant stiff axion field as the source of the birefringence angle for the CMB photons. 
We use the CB angle estimated by WMAP + Planck and ACT data: $\beta_{\rm CMB} = (0.264 \pm 0.058) \, \rm deg$~\cite{Diego-Palazuelos:2025dmh}.\\
Starting from the Lagrangian in Eq.~(\ref{eq:2.1}), the birefringence angle can be computed to be~\cite{Carroll:1989vb, Harari:1992ea}
\begin{equation}
    \beta(z)=\frac{g_{\phi\gamma}}{2}\left( \phi_0-\phi(z) \right) = \frac{g_{\phi\gamma}}{2}\Delta \phi ,
    \label{eq:CBphi}
\end{equation}
where $\phi_0$ is the field value today and $\phi(z)$ the field value at the point of the photon emission. In the comparison with data, we take the value at recombination $z=z_*$, i.e. $\phi(z_*)\equiv\phi_*.$
In the following, we consider a homogeneous scalar field.
We checked that field fluctuations can be neglected for the calculation of the birefringence.
From Eq.~(\ref{eq:2.8}), we can then determine the field displacement
\begin{equation}\label{eq:2.11}
    \phi_0-\phi_*=\int_{t_*}^{t_0}\dot{\phi}(t) \ dt= \int_{0}^{z_*} \frac{\dot{\phi}(z')}{(1+z')H(z')} \ dz',
\end{equation}
where $H(z)$ can be computed from concordance cosmology since this extra component is negligible at the epoch of recombination (as we will discuss in the next Section). The numerical result of Eq.~(\ref{eq:2.11}) is
\begin{equation}\label{eq:2.13}
    \phi_0-\phi_* \simeq \left( 2.1\times 10^{28} \ \text{eV}^{-1} \right)  \dot{\phi}_*\equiv t_A \dot{\phi}_*.
\end{equation}
Combining Eqs.~(\ref{eq:CBphi}) and~(\ref{eq:2.13}), 
we can now relate the axion-photon coupling to the time derivative of the field at the epoch of recombination
\begin{equation}\label{eq:2.14}
    g_{\phi\gamma}=\left( \frac{2\beta_{\rm CMB}}{t_A} \right) \frac{1}{\dot{\phi}_*}.
\end{equation}

\subsubsection{Birefringence angle of reionization photons} \label{Ssec:ReionizationBirefringence}
The redshift dependence of cosmic birefringence probes the evolution of the underlying field, offering a way to discriminate between models that predict the same rotation angle at recombination \cite{Gasparotto:2022uqo,Galaverni:2023zhv,Naokawa:2025shr,Yin:2026qlh}.
In particular, one could consider the birefringence angle for the CMB polarization signal produced at reionization, $\beta_{\rm reio}$, rather than at recombination as discussed above.
In general, models of axion dark energy predict $\beta_{\rm reio} = O(\beta_{\rm CMB})$ because dark energy starts to deviate from a cosmological constant only at low redshifts~\cite{Minotti:2026ivu}. Axion dark matter, due to faster oscillations, can imprint a more distinguishable signal~\cite{Galaverni:2023zhv}. On the other hand, the region in the ALPs parameter space preferred by cosmic birefringence is strongly constrained by the Lyman-alpha forest bound $m>2\times10^{-20} \ \text{eV}$~\cite{Rogers:2020ltq} and thus axion dark matter faces serious difficulties to explain the observed birefringence angle of CMB photons~\cite{Zhang:2024dmi}.

\begin{figure}[t!]
    \centering
\includegraphics[width=0.49\textwidth]{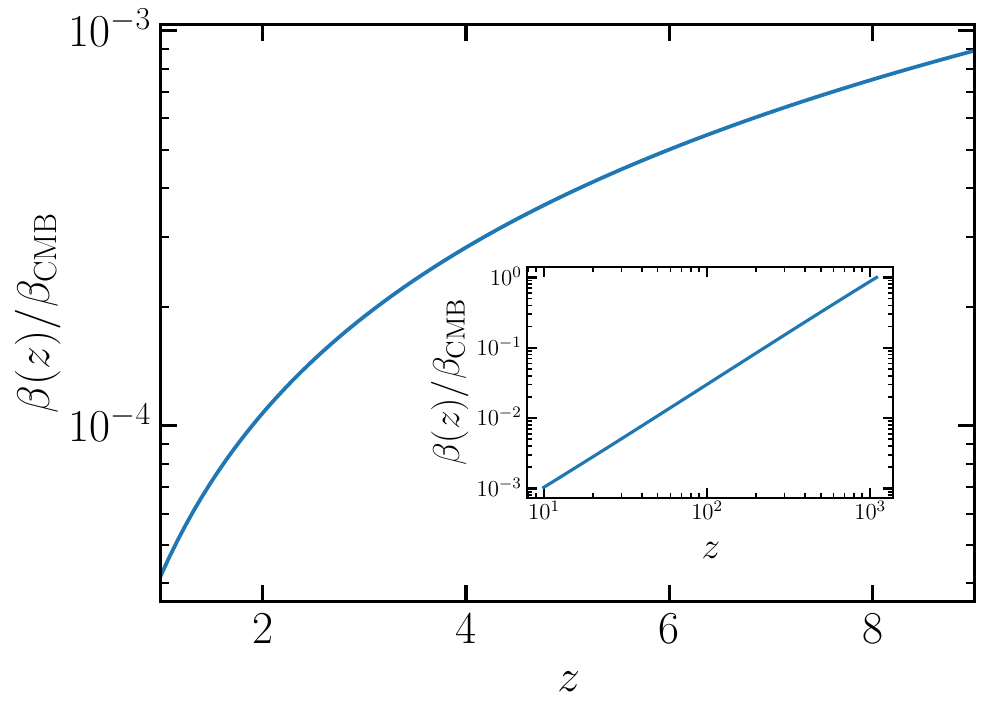}
    \centering    
    \caption{Birefringence angle as a function of redshift, normalized by the CMB birefringence angle. The main plot shows the behavior of the birefringence angle of reionization photons, whereas the inset extends the redshift domain up to recombination.}
    \label{fig:beta}
\end{figure}

In our stiff axion model, the ratio between the birefringence angles at a given redshift and at recombination is simply given by
\begin{equation}
\label{eq:beta(z)}
    \frac{\beta(z)}{\beta_{\rm CMB}}= \frac{\phi_0 -\phi(z)}{\phi_0 -\phi_*}=\frac{\int^{z}_0 \frac{(1+z')^2}{H(z')} \ dz'}{\int^{z_*}_0 \frac{(1+z')^2}{H(z')} \ dz'}.
\end{equation}
The results is plotted in Fig.~\ref{fig:beta}.\\
Using Eq.~(\ref{eq:beta(z)}), we predict a reionization birefringence angle $\beta_{\rm reio}/\beta_{\rm CMB}\approx 7.5\times 10^{-4}$.
Should the birefringence angle of reionization photons be consistent with zero, this  would establish a preference for a stiff pseudo-scalar explanation over the axion DE interpretation.\\

\subsection{Constraints from BBN}\label{Ssec:BBN}
The presence of a cosmological stiff-fluid component can affect the primordial abundances of elements predicted by BBN~\cite{Dutta:2010cu}.
We estimate an upper limit on the energy density of such component using Ref.~\cite{Zhang:2025vvz}, where
the primordial Helium-4 mass fraction $Y_p$ is related to the density parameter of the stiff fluid $K_{10}\equiv (\rho_{\phi}/\rho_{\gamma})_{T=10 \, \text{MeV}}$. Using the $2\sigma$ upper bound on the BBN Helium abundance $Y_p<0.2484$ 
from Ref.~\cite{Aver:2026dxv}, we find $K_{10}<28.5$.
We can then derive an upper bound on $\dot{\phi}_*$ 
from the energy densities of the axion field and photons at $T_{10}\equiv 10 \ \text{MeV}$:
\begin{equation}\label{eq:2.16}
    \frac{\dot{\phi}^2_*}{2}\left( \frac{T_{10}/T_0}{1+z_*}\right)^6\left[ \frac{g_{*,s}(T_{10})}{g_{*,s}(T_{0})}\right]^2<28.5 \left[ \frac{\pi^2}{15} T_{10}^4 \right].
\end{equation}
Using Eq.~(\ref{eq:2.14}) to express $\dot{\phi}_*$ in terms of $\beta$ and $g_{\phi\gamma}$, and taking
the 1$\sigma$ lower limit
$\beta_{\rm min}\equiv\beta_{\rm CMB,1\sigma}^{\rm low}=0.206 \, \text{deg}$,
Eq.~(\ref{eq:2.16}) leads to a lower bound on the axion-photon coupling in our stiff fluid scenario
\begin{equation}
    g_{\phi\gamma}>8.9\times10^{-14} \ \text{GeV}^{-1} \;.
    \label{eq:gBBN}
\end{equation}

\subsection{Constraint from axion-photon conversion in
astrophysical environments}\label{Ssec:conv}

Currently, the strongest upper bound on $ g_{\phi\gamma}$ at very low masses, that does not rely on the axion being dark matter, comes from a search for axion-photon conversion using the quasar H1821+643~\cite{Reynes:2021bpe}. At $95.5\%$ CL, it is
\begin{equation}
    g_{\phi\gamma}<5.0\times10^{-13} \ \text{GeV}^{-1} \ \ \ \  \ \ ( m_{\phi}<10^{-12} \ \text{eV}).
    \label{eq:gquas}
\end{equation}
Combining Eq.~(\ref{eq:gquas}) with Eq.~(\ref{eq:gBBN}), we therefore identify a narrow range for the axion-photon coupling that allows our stiff field to explain cosmic birefringence without affecting BBN or violating current astrophysical constraints.

\subsection{End of kination}\label{sec:EndKination}
\begin{figure}[t!]
    \centering
\includegraphics[width=0.49\textwidth]{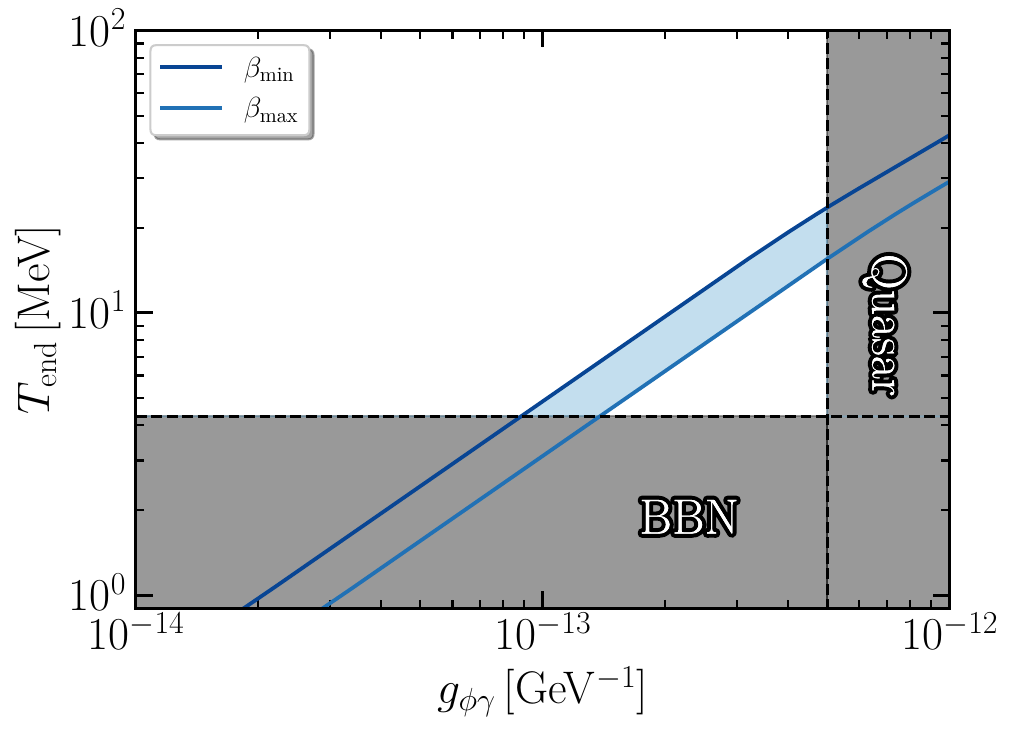}
    \centering    \caption{Temperature of the primordial plasma at the end of kination as a function of the axion-photon coupling constant. Constraints from BBN~\cite{Aver:2026dxv} and quasars~\cite{Reynes:2021bpe} are reported with gray regions, while the parameter space explaining cosmic birefringence data~\cite{Diego-Palazuelos:2025dmh} is in light blue. }
    \label{fig:T}
\end{figure}

The energy density of a stiff fluid scales as $(1+z)^{6}$, which implies an epoch of kination before the usual radiation era in the early Universe. The end of the kination epoch can be easily computed by equating the energy density of the axion field to that of relativistic particles
\begin{equation}\label{eq:2.19}
    \frac{\dot{\phi}^2_*}{2}\left( \frac{T_{\rm end}/T_0}{1+z_*} \right)^6\left[ \frac{g_{*,s}(T_{\rm end})}{g_{*,s}(T_{0})}\right]^2=\frac{\pi^2}{30}g_{*}(T_{\rm end})T_{\rm end}^4.
\end{equation}
Combining Eqs.~(\ref{eq:2.14}) and~(\ref{eq:2.19}), we can find the temperature of the primordial plasma at which the kination epoch ends
for a given value of the coupling constant $g_{\phi\gamma}$. The results are shown in Fig.~\ref{fig:T} with blue lines. The blue band encodes the 1$\sigma$ observational uncertainty on $\beta_{\rm CMB}$ from the measurement in Ref.~\cite{Diego-Palazuelos:2025dmh}, used in our analysis.
The allowed range for the end of kination is
\begin{equation}
\label{eq:Tend}
    4.3 \, \text{MeV} < T_{\rm end} < 23.6 \, \text{MeV},
\end{equation}
corresponding to
\begin{equation}
    2.6 \times 10^{10} < z_{\rm end} < 1.4 \times 10^{11},
    \label{eq:kinend}
\end{equation}
where the lower bound is set by the BBN constraints and the upper bound from the quasar limit, as illustrated in Fig.~\ref{fig:T}.

\subsection{Constraints on the duration of kination from $ \Delta N_{\rm eff}$}
\label{Ssec:dur}
Curvature perturbations of the metric generate scalar fluctuations of the kinaton field that behave as a radiation component~\cite{Eroncel:2025bcb}.
Its energy density is constrained by the cosmological bounds on the effective number of neutrino species, which therefore can be translated into a constraint on the duration of the kination era.

Following Ref.~\cite{Eroncel:2025bcb}:
\begin{equation}
\begin{aligned}
     \Delta N_{\rm eff}&=\frac{8}{7}\left( \frac{11}{4} \right)^{4/3} \left[\frac{g_*(T_{\rm end})}{g_{\gamma}} \right]\left[\frac{g_{*,s}(T_{0})}{g_{*,s}(T_{\rm end})} \right]^{4/3} \Omega_{\delta\phi}(k_{\rm end})\\
     &=6.21 \ \Omega_{\delta\phi}(k_{\rm end}),
\end{aligned}
\label{eq:deltaNeff}
\end{equation}
where $g_{*} $ and $g_{*,s}$ are the number of relativistic degrees of freedom in the energy and entropy densities, $g_{\gamma}=2  $ is the number of degrees of freedom of the photon and $T_{\rm end}$ and $T_0$ are the temperatures of the primordial plasma at the end of kination and today, respectively.
We take $g_*(T_{\rm end})=g_{*,s}(T_{\rm end})=10.75$ (justified by our range for $T_{\rm end}$).

The energy density of fluctuations over the background kination density at the end of the kination phase is~\cite{Eroncel:2025bcb}
\bea
    \Omega_{\delta\phi}(k_{\rm end})&=&\frac{4}{\pi\,k_{\rm end}}\int_{k_{\rm end}}^{k_{\rm kin}}dk \  \mathcal{P}_{\zeta}(k)\nonumber\\
    &=&\frac{4\,A_s}{\pi\,k_0^{n_s-1}n_s}\left(\frac{k_{\rm kin}^{n_s}}{k_{\rm end}}-k_{\rm end}^{n_s-1}\right),\label{eq:omegadeltaphi}
\eea
where we have used the primordial power spectrum of curvature perturbations $\mathcal{P}_{\zeta}(k)=A_s\,(k/k_0)^{n_s-1}$ with
$A_s\simeq2.099(29)\times 10^{-9}$, $n_s\simeq 0.965(4)$ and $k_0=0.05 \ \rm Mpc^{-1}$.
Here, $k_{\rm kin}$ and $k_{\rm end}$ are the modes crossing the Hubble radius at the beginning and end of the kination phase, respectively.
Since the power spectrum is nearly scale-invariant, it is clear from Eq.~(\ref{eq:omegadeltaphi}) that $\Omega_{\delta\phi}(k_{\rm end})$ essentially depends on the ratio $k_{\rm kin}/k_{\rm end}$ but not on their specific values. 
This ratio can be related to the number of e-foldings $N_k$ of the kination phase:
\begin{equation}
\label{eq.kkinkend}
    \frac{k_{\rm kin}}{k_{\rm end}}=\frac{(a\,H)_{\rm kin}}{(a\,H)_{\rm end}}= e^{N_k} \sqrt{\frac{1+e^{2N_k}}{2}}\;,
\end{equation}
where the last step is derived using the Friedmann equation for the Hubble rate $H$ at kination-radiation equality and at the beginning of kination, and including the energy density of radiation and kinaton components.

\begin{figure}[t!]
    \centering
\includegraphics[width=0.49\textwidth]{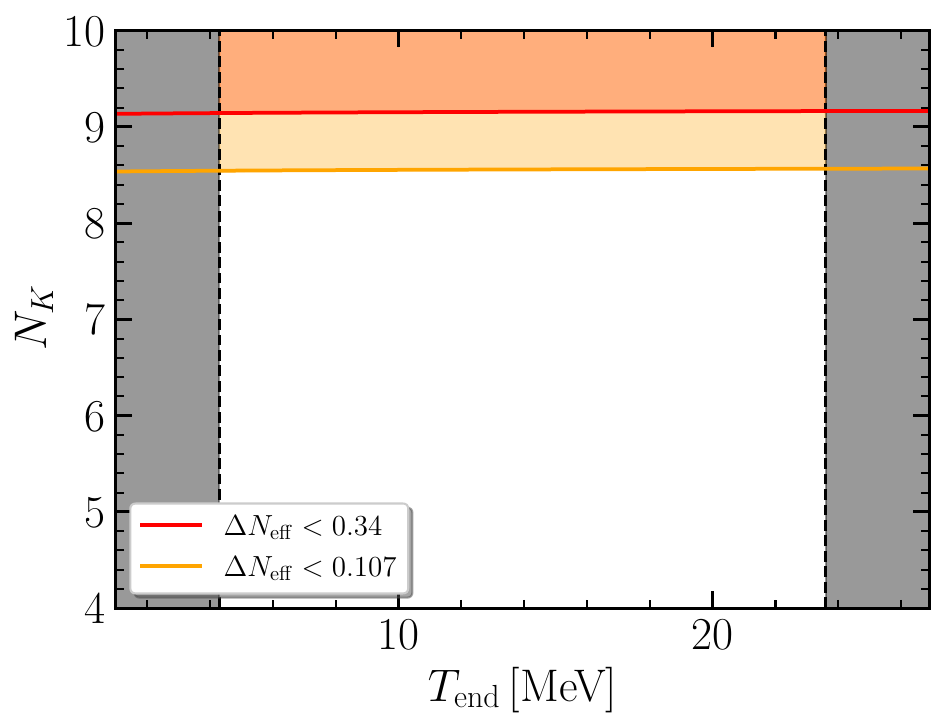}
    \centering    \caption{
    The orange (yellow) region shows the upper limit on number of e-foldings of kination, as a function of the temperature at the end of kination $T_{\rm end}$, from the constraint on the effective number of neutrino species $\Delta N_{\rm eff}<0.34$ ($\Delta N_{\rm eff}<0.107 $), as discussed in Sec.~\ref{Ssec:dur}. The allowed range of temperatures is obtained in Eq.~(\ref{eq:Tend}), see Fig.~\ref{fig:T}.}
    \label{fig:DurationKination}
\end{figure}

Concerning the experimental bound, we consider $\Delta N_{\rm eff}<0.34 \ (2\sigma)$ from the Planck experiment~\cite{Planck:2018vyg}, and a recent stronger limit obtained from the combination of BBN, CMB and BAO, $\Delta N_{\rm eff}<0.107 \ (95\% \ C.L.)$~\cite{Goldstein:2026iuu}.
The maximum number of e-foldings as a function of the temperature at the end of kination, is plotted in Fig.~\ref{fig:DurationKination}.
Note that it is nearly flat because of the nearly scale-invariant power spectrum, as anticipated above.

\subsection{Gravitational waves signals} \label{Ssec:GWs}

\begin{figure}[t!]
    \centering
    \includegraphics[width=0.42\textwidth]{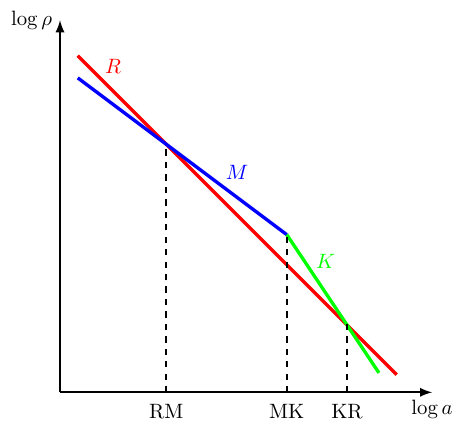}
\caption{Sketch of the evolution of the energy density of the Universe with early epochs of matter domination and kination
(R = radiation, M = matter, K = kination).}
\label{fig:sketch}
\end{figure}
We now discuss the phenomenology of the model in terms of gravitational wave (GW) signatures.
We focus on two possibilities: GWs induced by scalar curvature perturbations and the enhancement of primordial tensor modes from a  kination phase. 

First, we explore a cosmological scenario in which the kination phase is preceded by an epoch of early matter domination (MD), as it can occur in rotating axion models~\cite{Co:2019wyp,Co:2021lkc,Gouttenoire:2021wzu}.
If the transition between the matter and kination phases is sudden, scalar-induced GWs can be largely enhanced, realizing the so-called poltergeist mechanism~\cite{Harigaya:2023mhl,Inomata:2025wiv}~\footnote{See Ref. ~\cite{Bodas:2025wef} for another scenario producing large GW signals in rotating axion models.}. The evolution of the energy density of the Universe in this scenario is sketched in Fig.~\ref{fig:sketch}.

Following Ref.~\cite{Inomata:2025wiv} and in particular their Eqs. (C17) and (C18), inserting the ingredients of our model and after a bit of algebra, the GW density at a time $\bar{\eta}$ at the beginning of the the radiation era, after the end of kination, can be written as:
\bea
\label{eq:gwfinal}
&&    \Omega_{\rm GW}(\bar{\eta},k) \simeq
     \frac{3 \cdot 2^{-2n_s-8-\frac{1}{2}}}{625(3+2n_s)}
     A_s^2\left( \frac{k}{k_0} \right)^{2(n_s-1)}\\
     &&\times\left( 2\frac{k_*}{k} -1 \right)^{3+2n_s} \frac{k^6\eta_{\rm kin}^5(1+z_{\rm end})}{H_{\rm end}}\nonumber
     \\ 
     &&\times \left\{ \begin{array}{cl}
2.14 \ \frac{k \eta_{\rm kin}}{4} & : \ y_{\rm kin} \ll 1\\
1 & : \ y_{\rm kin} \gg 1 \ {\rm and} \ k \ll k_{*}.
\end{array} \right.\nonumber
\eea
$\eta_{\rm kin}$ is the conformal time at the beginning of the kination phase (matter-kination equality) and $\eta_{\rm RM}$ is the conformal time at the onset of the early MD phase (the first radiation-matter equality). We have defined $y_{\rm kin}=k\,\eta_{\rm kin}/4,$ while $z_{\rm end}$ and $H_{\rm end}$ are the redshift and Hubble rate at the end of the kination phase, respectively.
Here, we are considering the same (nearly) scale-invariant power spectrum of curvature perturbations discussed before.
The scale $k_*$ is defined as 
\begin{equation}
k_*=\min \left[k_{\rm max},\frac{1}{\eta_{\rm RM}}\right] 
\end{equation}
and $k_{\rm max}$ is an UV cut-off scale introduced to avoid non-linearities of the density perturbations, and for which we adopt the conservative choice of Ref.~\cite{Harigaya:2023mhl}.

\begin{figure}[!htbp]
    \centering
    \includegraphics[width=0.49\textwidth]{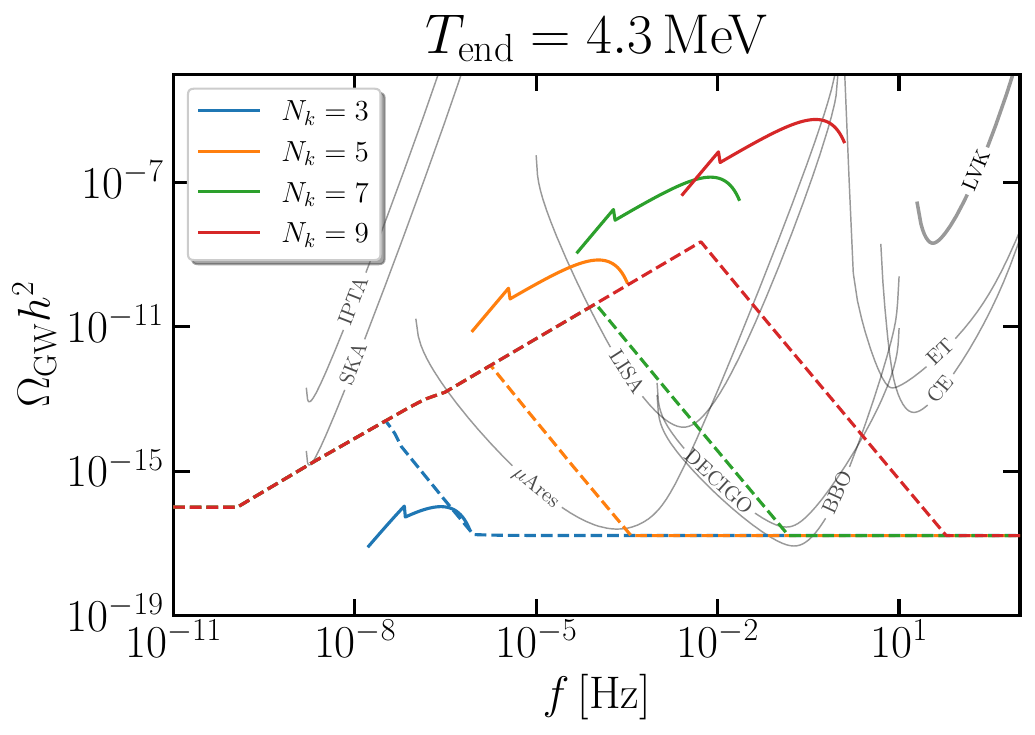}
    \par\vspace{1mm}
    \includegraphics[width=0.49\textwidth]{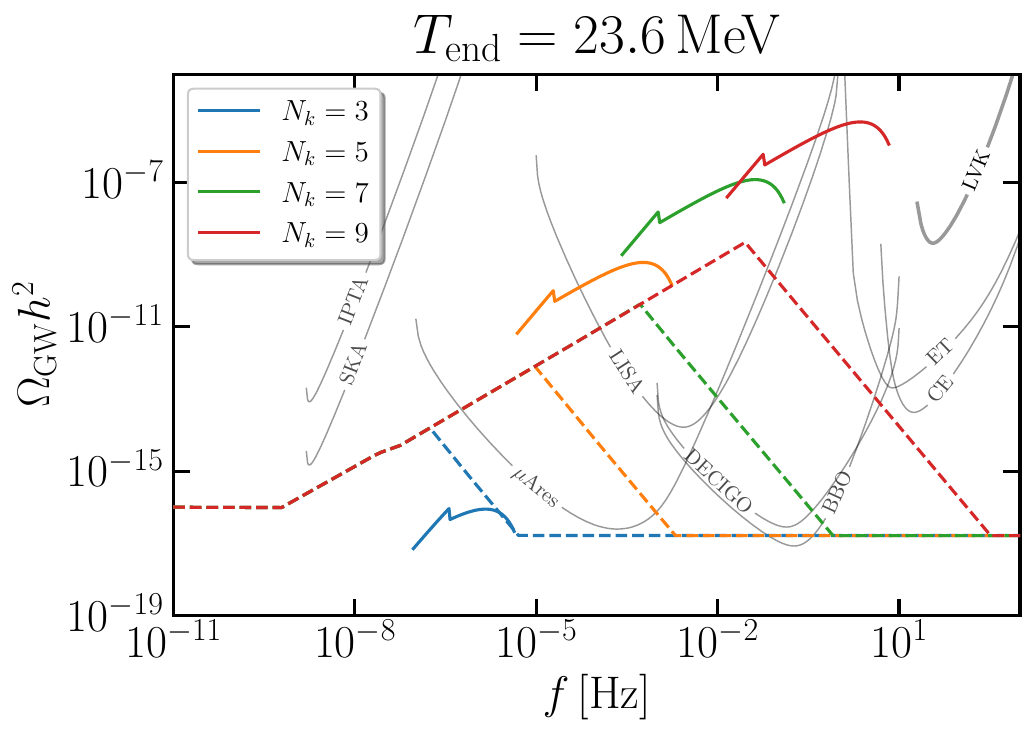}
    \caption{GWs produced by a kination phase via the poltergeist mechanism (solid lines) and via the amplification of the GWs background produced by inflation (dashed lines). We consider different values for the duration of the kination, expressed in terms of number of e-foldings $N_k$, and the temperature at the end of kination, $T_{\rm end}=4.3\,\mathrm{MeV}$ (\textbf{top panel}) and $T_{\rm end}=23.6\,\mathrm{MeV}$ (\textbf{bottom panel}). The thin black lines represent the power-law-integrated sensitivities of proposed GW experiments~\cite{Schmitz:2020syl,Sesana:2019vho,RoperPol:2021xnd}, normalized to $1\,\mathrm{yr}$ of observation time and $\mathrm{SNR}=1$. The thick black line shows the $2\sigma$ power-law-integrated sensitivity derived from LVK O1--O4a observations~\cite{Virgo:2025aai}. 
    }
    \label{fig:4.3}
\end{figure}

Finally, the GW energy density today is:
\begin{equation}
    \Omega_{\rm GW}(\eta_0,k) h^2 \simeq 0.39\left(  \frac{g_{*c}}{106.75}\right)^{-\frac{1}{3}}\Omega_{r0}h^2 \Omega_{\rm GW}(\bar{\eta},k)
    \label{eq:omegatoday}
\end{equation}
with $g_{*c}$ the effective relativistic degrees of freedom at a time $\eta_c$ during the kination phase when the GW-radiation energy ratio approaches a constant.

For a given choice of $z_{\rm end}$ (i.e., $T_{\rm end}$) and $N_K$, we can compute $\eta_{\rm kin}$ and then $\Omega_{\rm GW}(\eta_0,k)h^2$ from Eqs.~(\ref{eq:gwfinal}) and (\ref{eq:omegatoday}). 
Our predictions are shown in Fig.~\ref{fig:4.3} (solid colored lines) for the range of frequencies where linear perturbation theory is valid and the analytical expressions for $\Omega_{\rm GW}$ can be trusted. For a sufficiently long kination phase the GW signal is within the reach of proposed GW experiments.

Another potential GW signal arises from the amplification, during kination, of the stochastic GW background from inflationary tensor perturbations.
Following Refs.~\cite{Co:2021lkc,Gouttenoire:2021jhk}, the present-day GW energy density is given by
\bea
&&    \Omega_{\rm GW}(\eta_0,f)h^2 \simeq (1.4\times 10^{-17})\left( \frac{g_{*s}(T_{\rm hc})}{106.75} \right)^{-\frac{1}{3}}\\
     && \times \left( \frac{E_{\rm inf}}{10^{16} \rm \ GeV} \right)^4 \times \left\{ \begin{array}{cl}
1 & : f<f_{\rm end}\\
\frac{f}{f_{\rm end}}  & :  f_{\rm end}<f<f_{\rm kin}\\
\frac{f_{\rm kin}}{f_{\rm end}} \left( \frac{f_{\rm kin}}{f} \right)^2  & :  f_{\rm kin}<f<f_{\rm RM}\\
\frac{f_{\rm kin}}{f_{\rm end}} \left( \frac{f_{\rm kin}}{f_{\rm RM}} \right)^2  & :  f_{\rm RM}<f
\end{array} \right.\nonumber
\eea
where $T_{\rm hc}$ is the temperature when the mode $k=2\pi f$ crosses the Hubble radius, $f_{\rm kin}$ is related to $f_{\rm end}$ by Eq.~\ref{eq.kkinkend} and $f_{\rm RM}=f_{\rm end}e^{3N_k}$.
The inflationary energy scale $E_{\rm inf}$ sets the amplitude of the primordial tensor power spectrum. We take $E_{\rm inf}=1.35\times 10^{16} \, \mathrm{GeV}$, around the upper limit for standard single-field slow-roll inflation derived from Ref.~\cite{Tristram:2021tvh}.

The energy density of the amplified GW background sourced by inflation is plotted in Fig.~\ref{fig:4.3} (dashed colored lines).
For a sufficiently long duration of the kination epoch, the predicted GW signal is within the sensitivity windows of future experiments.

\subsection{PTA Gravitational waves}
\label{Ssec:PTA}

\begin{figure}[]
    \centering
    \includegraphics[width=0.49\textwidth]{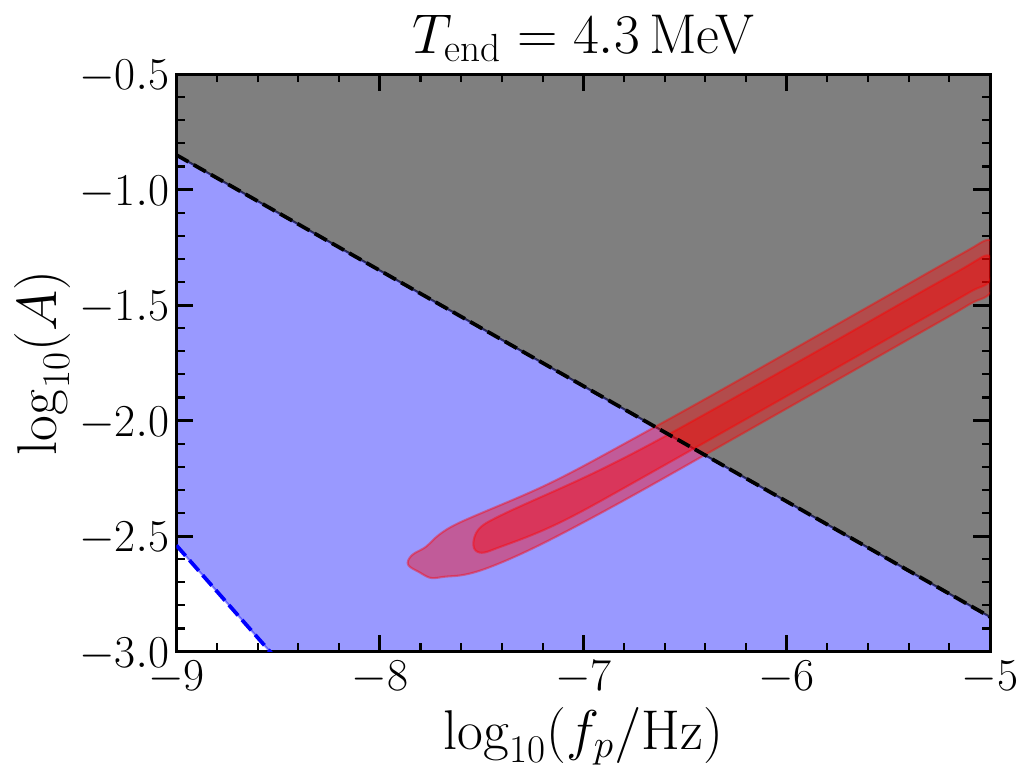}
    \hfill
    \includegraphics[width=0.49\textwidth]{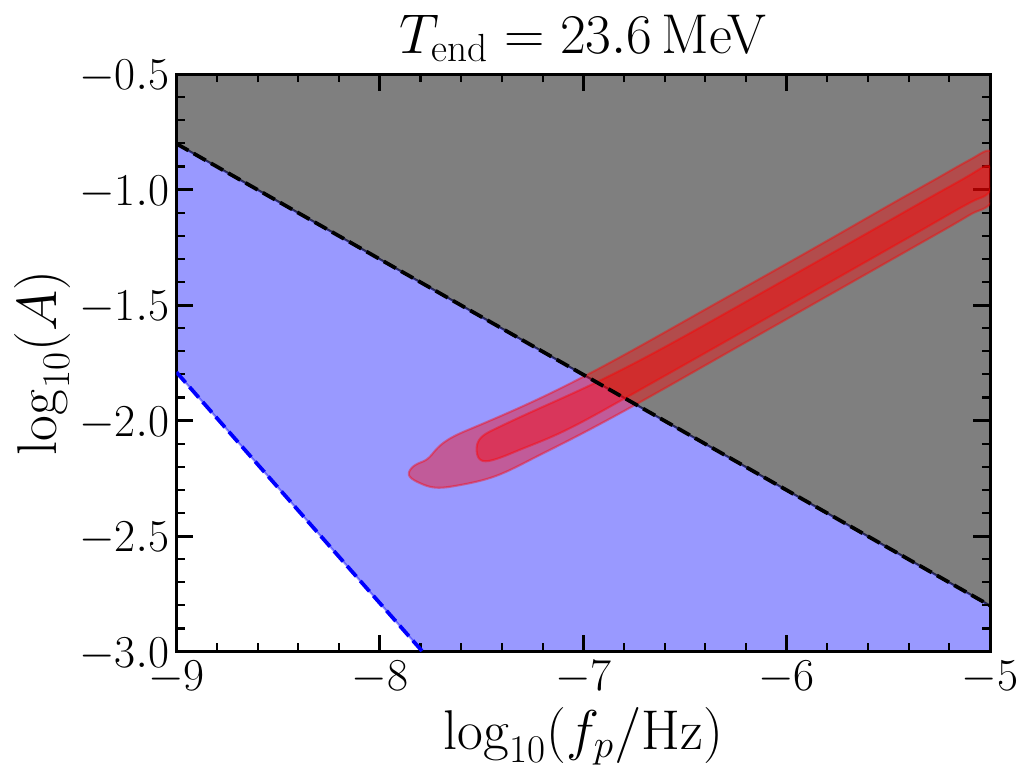}
    \caption{Red contours show the posteriors for the amplitude $A$ and peak-frequency $f_p$ of the power spectrum of curvature perturbations in Eq.~\ref{Spectrum} from the analysis of NANOGrav 15-year data~\cite{NANOGrav:2023gor,NANOGrav:2023hvm}.
    We consider $T_{\rm end}=4.3 \ \rm MeV$ (upper panel) and $T_{\rm end}=23.6 \ \rm MeV$ (lower panel).
    Shaded regions are excluded by $\Delta N_{\rm eff}$ constraints on the energy density in gravitational waves (gray) and in kinaton field fluctuations (blue).
    }
    \label{fig:NEW_PTA}
\end{figure}

In 2023, the Pulsar Timing Array (PTA) collaborations reported compelling evidence of a stochastic gravitational-wave background (SGWB) in the nano-hertz frequency range, see Ref.~\cite{InternationalPulsarTimingArray:2023mzf} for a comparison of the results obtained by different PTA groups.
Proposed explanations include a population of supermassive black hole binaries, as well as cosmological sources such as inflation, first-order phase transitions, cosmic strings, domain walls, and scalar-induced gravitational waves (GWs)~\cite{NANOGrav:2023hvm,EPTA:2023xxk}.

Scalar-induced GWs, i.e. GWs induced from curvature perturbations, can accommodate the observed amplitude and spectral index $n_T=d\log\Omega_{\rm GW}/d\log f \simeq 2$ of the signal, provided that the primordial curvature perturbations are enhanced at small scales. However, such large scalar perturbations can also produce a sizeable abundance of primordial black holes, potentially leading to strong constraints on this interpretation~\cite{Franciolini:2023pbf,Cai:2023dls,Figueroa:2023zhu}.
In Ref.~\cite{Harigaya:2023pmw}, it was shown that these bounds can be evaded if the scalar-induced GWs are generated during an epoch of kination, yielding a viable scenario for the PTA signal.
In this subsection, we revisit this possibility in the context of the kination era in Eq.~\ref{eq:kinend}.

Following Ref.~\cite{Harigaya:2023pmw}, we model the small-scale enhancement of the curvature power spectrum with a log-normal parametrization:
\begin{equation}
\label{Spectrum}
    \mathcal{P}_{\zeta}(k)=\frac{A}{\sqrt{2\pi\Delta^2}}\exp\left[ -\frac{\ln^2({k/k_p})}{2\Delta^2} \right],
\end{equation}
where $A$ and $\Delta$ are the amplitude and width of the peak, and $k_p$ is the peak wave number. We take $\Delta=1.$
One finds that for $\Delta\gtrsim 1$, the density of scalar-induced GWs produced during the kination era scales as $\Omega_{\rm GW}\sim A^2\,f^2/(f_{\rm end}\,f_p)$, i.e., it exhibits the $f^2$ tail required to fit the PTA data. This behavior persists for other non-sharply-peaked choices of the power-spectrum~\cite{Harigaya:2023pmw}.
In the following, we focus on the NANOGrav 15-year data~\cite{NANOGrav:2023gor,NANOGrav:2023hvm}.
We consider the 14 frequency HD-correlated free spectrum from Ref.~\cite{the_nanograv_collaboration_2026_21844115} and analyze it using the PTArcade Ceyffl wrapper \cite{andrea_mitridate_2023_8106173,william_g_lamb_2023_7775774}.
In Fig.~\ref{fig:NEW_PTA} we show the joint posterior of the amplitude and peak scale of the curvature power spectrum. We consider two values of the temperature at the end of kination, $T_{\rm end}=4.3 \ \rm MeV$ and $T_{\rm end}=23.6 \ \rm MeV$, which bracket our allowed range in Eq.~\ref{eq:Tend}. 
Our contours agree with the findings of Ref.~\cite{Harigaya:2023pmw}.
As discussed in Ref.~\cite{Harigaya:2023pmw}, the total energy density in GWs is constrained by the requirement that they do not overproduce dark radiation.
This bound excludes the gray region in Fig.~\ref{fig:NEW_PTA}.
However, an additional constraint, not considered in Ref.~\cite{Harigaya:2023pmw}, arises from the fluctuations of the field driving the kination era.
As discussed in Sec.~\ref{Ssec:dur}, these perturbations are sourced during the kination phase and their energy density redshifts as radiation, therefore contributing to the effective number of neutrino species. 
We derive a limit using $\Delta N_{\rm eff}=6.21 \, \Omega_{\delta\phi}(k_{\rm end})$ from Eq.~\ref{eq:deltaNeff}, and computing $\Omega_{\delta\phi}(k_{\rm end})$ from the first line of Eq.~\ref{eq:omegadeltaphi} with the primordial power spectrum given by Eq.~\ref{Spectrum}.
By imposing $\Delta N_{\rm eff}<0.34$~\cite{Planck:2018vyg} we exclude the blue regions in Fig.~\ref{fig:NEW_PTA}.
As the figure shows, the interpretation of the PTA data in terms of scalar-induced GWs produced during an epoch of kination is disfavored.

\section{Conclusions}\label{Sec:Conclusion}
In this work, we have studied a new explanation for the measured birefringence angle of CMB photons~\cite{Diego-Palazuelos:2025dmh}
in terms of an axion-like scalar field behaving as a stiff fluid ($w=1$). Throughout the epochs relevant for birefringence, from recombination to the present day, this field remains a subdominant component of the total energy density of the Universe.
At earlier times, before BBN, the field dominated the energy density and drove a phase of kination.
Requiring consistency with BBN predictions, and accounting for the measured birefringence angle, we derived a lower bound on the axion-photon coupling of $g_{\phi\gamma}>8.9\times 10^{-14} \ \text{GeV}^{-1}$. 
The most stringent upper limit on this coupling comes from the non-observation of photon-axion conversion in the quasar H1821+643~\cite{Reynes:2021bpe}, $g_{\phi\gamma}<5.0\times10^{-13} \ \text{GeV}^{-1}$. Therefore, the allowed range of the axion-photon coupling is narrow and will be probed in the coming years with astrophysical axion searches. 

We have shown that this model predicts a suppressed birefringence angle from reionization. Therefore, probes of the cosmic birefringence at different redshifts can discriminate between the stiff-axion and evolving-dark-energy interpretations.

We have discussed the gravitational-wave phenomenology of the model considering two possible signals.
If kination is preceded by an early matter-dominated epoch, the sudden transition between the two phases can enhance scalar-induced GWs~\cite{Inomata:2025wiv}. In addition, kination amplifies the irreducible stochastic GW background from inflationary tensor perturbations~\cite{Co:2021lkc,Gouttenoire:2021jhk}.
We have shown that, in the range of kination temperatures for which the model explains the observed CMB birefringence, these GWs signals can be within reach of future GW experiments.
Finally, we have revisited the interpretation of the stochastic GW background reported by the PTA collaborations in terms of scalar-induced GWs generated during a kination phase, and found that this scenario is disfavored by dark radiation bounds.

\vspace{1cm}

{\it Note added}: During the completion of this work, Ref.~\cite{Co:2026rga} appeared on arXiv. Ref.~\cite{Co:2026rga} focuses on the specific model of rotating axion, while the goal of our work is instead to provide a more general framework to explain CMB birefringence with a stiff fluid.
\vspace{1cm}

\section*{Acknowledgements}
We acknowledge support from the research grant TAsP {\sl TAsP (Theoretical Astroparticle Physics)} funded by Istituto Nazionale di Fisica Nucleare (INFN).
The work of NF, AM, MR, YW and TZ is supported by the European Union – Next Generation EU and by the Italian Ministry of University and Research (MUR) via the PRIN 2022 project n. 20228WHTYC - CUP D53C24003550006. TZ acknowledges support from the UniTO-IBS Avogadro Program (collaboration agreement between the University of Torino and the Institute for Basic Science, South Korea).

\appendix

\section{A scenario without kination}
\label{Sec:Nokination}

A different situation from what discussed in the main text arises if the pseudo-scalar field exhibits a stiff behavior from around recombination to the present day, but had a different equation of state ($w\neq1$) in the past, always remaining a subdominant component of the energy density of the universe.
For example, an ordinary matter ($w=0$) fluid could evolve into a stiff ($w=1$) fluid as depicted in Fig.~\ref{fig:sketch}, while its energy density remains below that of the radiation component at all times.

In such a scenario, there was never a kination era and, therefore, the constraint from BBN that we used in the main text no longer applies.
We can derive a conservative bound requiring the energy density of the scalar field to be subdominant at the CMB time. We find
\begin{equation}
    g_{\phi\gamma}>2.0\times10^{-21}\ \rm GeV^{-1}.
\end{equation}
This is the lowest coupling a stiff pseudo-scalar field must have in order to explain the measured CMB birefringence angle.
The dependence of the birefringence angle on the redshift in this alternative scenario is equivalent to what we plotted in Fig.~\ref{fig:beta} of the main text.

\bibliographystyle{bibi}
\bibliography{bibliography}

\end{document}